\documentclass[twocolumn, prb, showpacs]{revtex4-1}
\usepackage{graphicx}
\usepackage{dcolumn}
\usepackage{bm}
\usepackage{color}

\begin{document}

\title{Sodium layer chiral distribution and spin structure of \\Na$_2$Ni$_2$TeO$_6$ with a honeycomb network}

\author{Sunil K. Karna,$^{1}$ Y. Zhao,$^{2,3}$ R. Sankar,$^{1,4}$ M. Avdeev,$^{5}$ P. C. Tseng,$^{6}$ C. W. Wang,$^{7}$ G. J. Shu,$^{1}$ K. Matan,$^{8}$ G. Y. Guo,$^{6,9}$ and F. C. Chou$^{1,10,11}$}
\affiliation{$^1$Center for Condensed Matter Sciences, National Taiwan University, Taipei 10617, Taiwan}
\affiliation{$^2$NIST Center for Neutron Research, National Institute of Standards and Technology, Gaithersburg, Maryland 20899, USA}
\affiliation{$^3$Department of Materials Science and Engineering, University of Maryland, College Park, Maryland 20742, USA}
\affiliation{$^4$Institute of Physics, Academia Sinica, Taipei 10617, Taiwan}
\affiliation{$^5$Australian Nuclear Science and Technology Organisation, Locked Bag 2001, Kirrawee DC NSW 2232, Australia}
\affiliation{$^6$Department of Physics, National Taiwan University, Taipei 10617, Taiwan}
\affiliation{$^7$Neutron Group, National Synchrotron Radiation Research Center, Hsinchu 30076, Taiwan}
\affiliation{$^8$Department of Physics, Faculty of Science, Mahidol University, Bangkok 10400, Thailand}
\affiliation{$^9$Physics Division, National Center for Theoretical Sciences, Hsinchu 30013, Taiwan}
\affiliation{$^{10}$National Synchrotron Radiation Research Center, Hsinchu 30076, Taiwan}
\affiliation{$^{11}$Taiwan Consortium of Emergent Crystalline Materials, Ministry of Science and Technology, Taipei 10622, Taiwan}

\date{\today}

\begin{abstract}

The nature of Na ion distribution, diffusion path, and the spin structure of $P2$-type Na$_2$Ni$_2$TeO$_6$ with a Ni honeycomb network has been explored.  The nuclear density distribution of Na ions reveals a 2D chiral pattern within Na layers without breaking the original 3D crystal symmetry, which has been achieved uniquely via an inverse Fourier transform (iFT)-assisted neutron diffraction technique. The Na diffusion pathway described by the calculated iso-surface of Na ion bond valence sum (BVS) map is found consistent to a chiral diffusion mechanism. The Na site occupancy and Ni$^{2+}$ spin ordering were examined in detail with the electron density mapping, neutron diffraction,  magnetic susceptibility, specific heat, thermal conductivity and transport measurements.  Signatures of both strong incommensurate (ICM) and weak commensurate (CM) antiferromagnetic (AFM) spin ordering were identified in the polycrystalline sample studied, and the CM-AFM spin ordering was confirmed by using a single crystal sample through the $k$-scan in the momentum space corresponding to the AFM peak of ($\frac{1}{2}$, 0, 1).

\end{abstract}



\maketitle

\section{Introduction}

   The layered transition metal oxides have received special attention in condensed matter physics research because of their rich physical properties, including ion intercalation for potential battery applications, superconductivity, and intercalant-sensitive magnetic phase transitions.~\cite{Terasaki1997, Schaak2003, Lee2006, Shu2016, Weller2009, Berthelot2011, Medarde2013, Shu2013, Linden2010} The group IA ions that are sandwiched between the transition metal oxide layers act as passive charge reservoirs to influence the electronic structure.  $P2$-type Na$_2$Ni$_2$TeO$_6$ has been shown to exhibit high ionic conductivity at room temperature for potential applications as a separator in the Na-ion battery.~\cite{Berthelot2012, Evstigneeva2011,Gupta2013}   The mixed edge-sharing (Te/Ni)O$_6$ octahedra in each layer create a unique Te-centered NiO$_6$ honeycomb network, which leads to a complex potential field profile that is used to drive the Na ion diffusion in 2D.  
   
   X-ray diffraction studies have shown that the crystal structure of Na$_2$Ni$_2$TeO$_6$ can be indexed with a space group $P6_3/mcm$ satisfactorily, and a 3D antiferromagnetic (AFM) spin ordering of T$_N$$\sim$27 K has been proposed; however, the impact of Na ion diffusion on the crystal and spin structure has never been explored in detail, in contrast to its sister compound Na$_x$CoO$_2$ which has been studied extensively and shown rich physical insights.\cite{Lee2006, Shu2008, Shu2016, Weller2009, Berthelot2011, Medarde2013, Shu2013, Evstigneeva2011, Sankar2014}  It is highly desirable to investigate the 2D structure of the Na-layer that reflects the potential field generated by the (Te/Ni)-O honeycomb sublattice and its relationship to the magnetic structure of Ni spins. In particular, due to the diffusive nature of intercalated Na ions, a unique dynamic magneto-phonon coupling of the system could be revealed through the  detailed analysis of the Na ion distribution and crystal structure change as a function of temperature. Below we present an integrated synchrotron X-ray and high-resolution neutron powder diffraction study that clearly reveals an in-plane chiral circular pattern within the Na-layers. An inverse Fourier transform (iFT) technique was applied to provide real space information from the diffraction data for a more accurate Na ion site assignment on structure refinement. Comparing to the calculated bond valence sum (BVS) map of Na, the Na-ion diffusion in 2D is found to follow an effective translational path of alternating chirality change.\\

\begin{figure*}
\begin{center}
\includegraphics[width=5.5in]{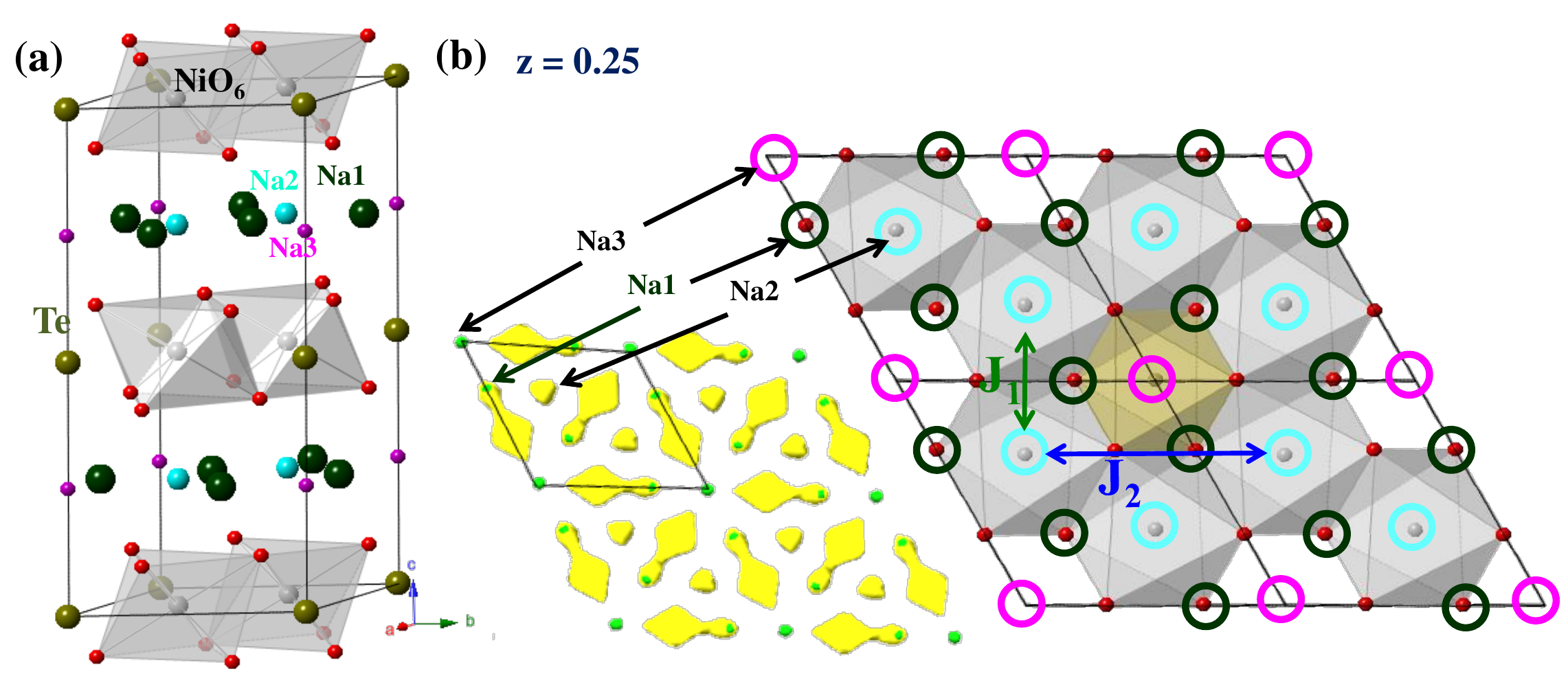}
\end{center}
\caption{\label{fig-crystru} (a) The crystal structure of Na$_2$Ni$_2$TeO$_6$ of space group $P6_3/mcm$ with three major Na sites (Na1, Na2, and Na3) assigned following Ref.~\onlinecite{Evstigneeva2011}.  (b) The 2D honeycomb network is composed of edge-shared NiO$_6$ with Te (also edge-shared TeO$_6$ octahedron) sitting in each honeycomb center, with Na sites in the neighboring layer shown as circles. The electron density map obtained from the iFT-assisted X-ray diffraction at z=0.25 is shown on the left. The effective magnetic exchange couplings $J_1$ and $J_2$ are indicated.}
\end{figure*}

\begin{figure}
\begin{center}
\includegraphics[width=3.5in]{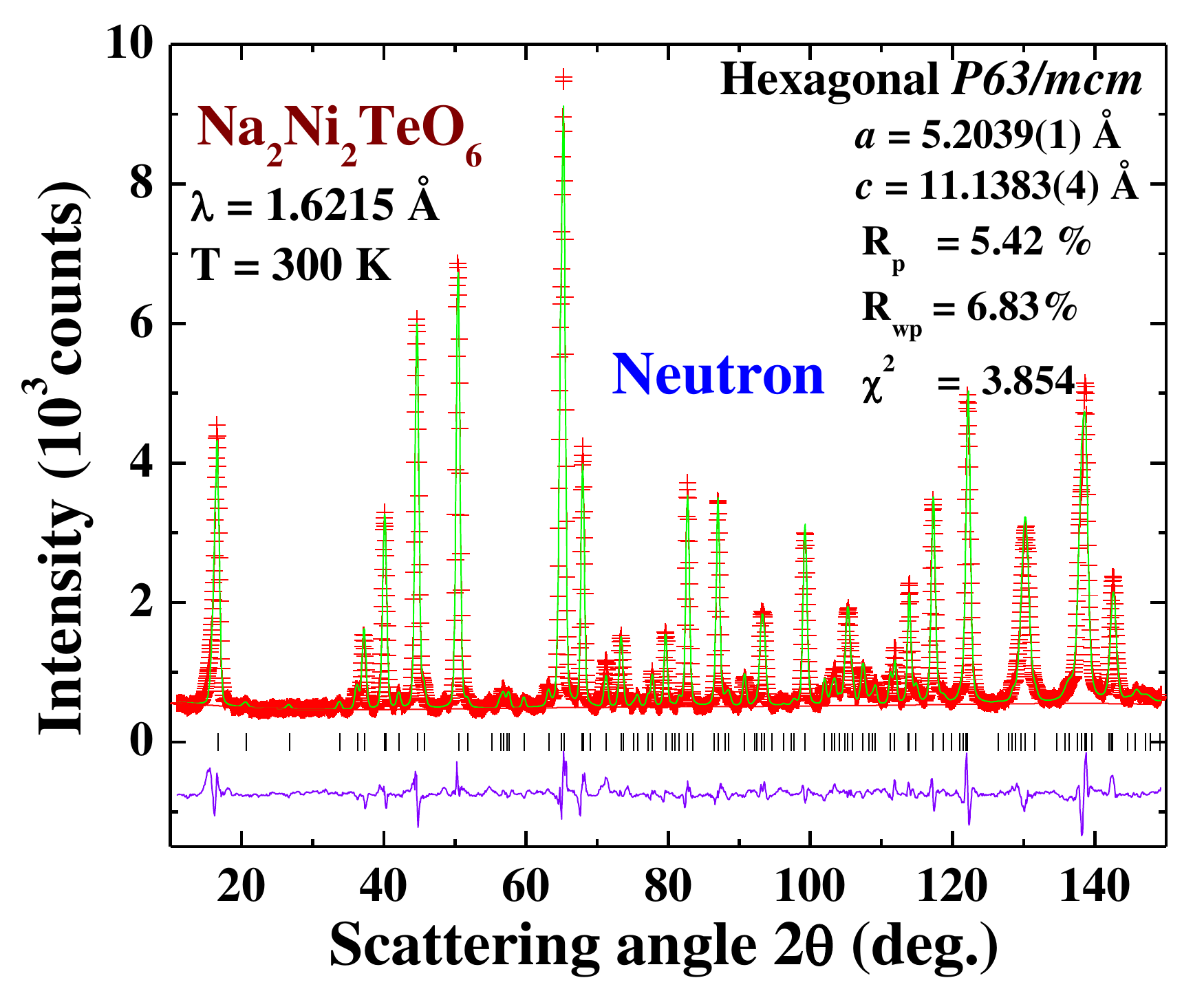}
\end{center}
\caption{\label{fig-NPD} Observed (crosses) and fitted (solid lines) high-resolution neutron powder-diffraction patterns taken at 300 K, assuming a Hexagonal crystal structure with the $P6_3/mcm$ space group. The differences between the calculated and observed patterns are plotted at the bottom. The solid vertical lines mark the calculated positions of the Bragg reflections of the proposed crystalline structure. } 
\end{figure}

\section{Experimental details and theoritical calculations}

 The experimental details of the polycrystalline sample preparation and single crystal growth of Na$_2$Ni$_2$TeO$_6$ have been reported in our previous work.~\cite{Sankar2014} The preliminary X-ray diffraction measurements were performed on a Bruker D8 ADVANCE diffractometer employing Cu K$_\alpha$ radiation, and Synchrotron X-ray was performed at NSRRC using BL01C2 beam line. Neutron powder diffraction data between 2 to 450 K were collected on the high-resolution neutron powder diffractometer (HRPD) ECHIDNA [monochromator: Ge (331), $\lambda = 2.439 $ \AA$ $] and the high-intensity powder diffractometer (HIPD) WOMBAT [monochromator: Ge (113), $\lambda = 2.9502  \AA $], installed at the OPAL reactor in ANSTO, Australia.  Approximately $\sim$7 grams of the powder sample was loaded into a cylindrical vanadium-can which gave rise to no measurable background diffraction peaks. Neutron scattering experiments on powder and single crystal samples were also conducted using the NG-5 triple-axis spectrometer SPINS at the NIST Center for Neutron Research [$\lambda=4.702 \AA$]. Cooled beryllium filter was used to reduce the higher-order neutron contaminations. The magnetic susceptibility measurement was performed using a SQUID-VSM magnetometer (Quantum Design USA).  Chemical analysis was performed using Electron Probe X-Ray Microanalyzer (EPMA) taken with a JEOL JXA-8200 analyzer. Bond valence sum (BVS) was calculated using the program 3DBVSMAPPER.\cite{Sale2012} The nuclear or electron density maps were obtained employing the General Structure Analysis System (GSAS) or FullProf programs.  Starting with the profile refinement of neutron or X-ray diffraction pattern, the electron density distribution can be obtained by calculating the inverse Fourier transform of the structure factors. The scattering density $\rho({\bf r}) = \frac{1}{V} \sum_{H}F_{obs}({\bf H}) exp\{-2\pi i(\bf H.\bf r)\}$ was calculated using the program $Fourier$ (GSAS) or $GFourier$ (FullProf), where V is the volume of the unit cell, H is a reciprocal space vector, \textbf{r} is a vector position inside the unit cell, and $F_{obs}({\bf H})  = \vert F({\bf H})\vert exp\{i\phi({\bf H})\}$, where $\vert F({\bf H})\vert$ is the square root of integrated intensities taken directly from the measurement, and the calculated phase $\phi({\bf H})$ derived from the structural model by the GSAS or FullProf program is used to perform the Fourier transform.

Theoretical calculations were performed based on first-principles density functional theory (DFT) with generalized gradient approximation (GGA).\cite{Perdew1996} To describe the electron-electron correlation associated with the 3d states of Ni, the GGA plus on-site coulomb repulsion (GGA +U) calculations are performed with an effective U$_{eff}$ = (U - J) = 6.0 eV.~\cite{Dudarev1998, Jain2011} We have used the frozen-core full-potential projector-augmented wave (PAW) method~\cite{Blochl1994,Kresse1999} as implemented in the Vienna ab initio simulation package (VASP).\cite{Kresse1993} The wave functions are expressed in a plane wave basis set with an energy cutoff of 400 eV, and the self-consistent field energies are converged up to 10$^{-5}$ eV. For our calculations, we have used four metastable magnetic configuraions to estimate the magnetic couplings $J_1$, $J_2$ and $J_3$ of Ni atoms of interatomic distances 3.00194, 5.19951 and 5.56422~\AA (the three shortest interatomic distances) respectively. To explore the magnetic ground state, a supercell of size 4 times the size of primitive cell is considered, and there are 16 Ni atoms in the unit cell. In the present calculations, we used the tetrahedron method with $Bl\ddot{o}chl$ corrections for the Brillouin zone integration with a $\Gamma$-centered Monkhorst-Pack k-point mesh of $(9\times 9\times 7)$. The magnetic couplings $J_1$, $J_2$ and $J_3$ among the Ni atoms are expressed in terms of the spin Heisenberg Hamiltonian, $H=E_0-\sum J_{ij} {\bf S_i\cdot S_j}$, where $J_{ij}$ is the exchange interaction parameter between the Ni atoms at site $i$ and site $j$, and ${\bf S_i}$ (${\bf S_j}$) is the unit vector that represents the direction of the local magnetic moment at site $i~(j)$.  For an antiferromagnetic interaction, $J<0$ is assumed, and for a ferromagnetic interaction, $J>0$ is assumed, and the constant E$_0$ contains all of the spin-independent interactions.

\begin{table}
\caption{List of the refined structural parameters of Na$_2$Ni$_2$TeO$_6$ at 300 K, based on the assigned Na positions similar to those reported in the literature by Evstigneeva \textit{et al.}, and the iFT-assisted refinement from this study (lower), where $B_{iso}$ represents the isotropic temperature parameter and M represents the multiplicity.   }
\centering
\begin{tabular}{c rrrrrr}
\hline\hline
    \multicolumn{7}{c}{\textbf{Na$_2$Ni$_2$TeO$_6$}} \\
     \multicolumn{7}{c}{Hexagonal $P6_3/mcm$ space group ( No. 193)} \\
      \multicolumn{7}{c}{T=300 K, a = b = 5.2036(3) $ \AA$, c = 11.1387(3)$ \AA$} \\  
\hline
 Atom         & x               & y               & z               & M     & $B_{iso}$ ($ \AA$$^2$)    & Occupancy\\
\hline
Te           &  0               & 0              &  0	            & 2a     & 0.19(3)                  & 1.000     \\
Ni           & $\frac{2}{3}$    & $\frac{1}{3}$  &  0               & 4c     & 0.30(2)                  & 0.997(3)  \\
O            & 0.3112(2)        & 0.3112(2)      &  0.5944(1)       & 12b    & 0.57(4)                  & 1.002(3)  \\
Na1          & 0.6306(3)        & 0.0            &  $\frac{1}{4}$   & 6g     & 4.98(5)                  & 0.508(4)  \\ 
Na2          & $\frac{1}{3}$    & $\frac{2}{3}$  &  $\frac{1}{4}$   & 4c     & 3.12(3)                  & 0.203(2)  \\
Na3          & 0                & 0.0            &  $\frac{1}{4}$   & 2a     & 2.88(3)                  & 0.073(1)  \\ \\[0.2ex]
\multicolumn{7}{c}{$\chi$$^2$ = 5.654, $R_p$ = 6.92\%, $R_{wp}$ = 8.81\% } \\
\hline\hline

    \multicolumn{7}{c}{\textbf{Na$_2$Ni$_2$TeO$_6$}} \\
     \multicolumn{7}{c}{Hexagonal $P6_3/mcm$ space group ( No. 193)} \\
      \multicolumn{7}{c}{T=300 K, a = b = 5.2039(1) $ \AA$, c = 11.1383(4)$ \AA$} \\  

\hline
 Atom         & x               & y               & z               & M     & $B_{iso}$ ($ \AA$$^2$)    & Occupancy\\
\hline
Te           &  0               & 0              &  0	            & 2a     & 0.18(5)                  & 1.000     \\
Ni           & $\frac{2}{3}$    & $\frac{1}{3}$  &  0               & 4c     & 0.28(4)                  & 1.003(2)  \\
O            & 0.3112(2)        & 0.3112(2)      &  0.5943(1)       & 12b    & 0.57(6)                  & 0.992(2)  \\
Na1a         & 0.3538(4)        & 0.0            &  $\frac{1}{4}$   & 6g     & 2.28(4)                  & 0.012(2)  \\ 
Na1b         & 0.415(3)        & 0.072            &  $\frac{1}{4}$   & 6g     & 2.28(4)                 & 0.051(3)  \\ 
Na1c         & 0.6250(5)        & 0.0            &  $\frac{1}{4}$   & 6g     & 6.74(7)                  & 0.038(2)  \\
Na1d         & 0.6921(5)        & 0.065            &  $\frac{1}{4}$   & 6g     & 6.74(7)                & 0.165(2)  \\
Na1e         & 0.6252(2)        & 0.1042(3)      &  $\frac{1}{4}$   & 12j    & 0.91(2)                  & 0.018(1)  \\
Na2a         & $\frac{2}{3}$    & $\frac{1}{3}$  &  $\frac{1}{4}$   & 4c     & 1.44(3)                  & 0.103(3)  \\
Na2b         & 0.5639(4)        & 0.1872(1)      &  $\frac{1}{4}$   & 12j    & 0.91(1)                  & 0.013(4)  \\ 
Na2c         & 0.6881(2)        & 0.2190(4)      &  $\frac{1}{4}$   & 12j    & 3.71(2)                  & 0.014(2)  \\
Na3a         & 0                & 0.0            &  $\frac{1}{4}$   & 2a     & 1.60(4)                  & 0.043(1)  \\
Na3b         & 0.8570(4)        & 0.0            &  $\frac{1}{4}$   & 6g      & 4.37(3)                 & 0.012(3)  \\ \\[0.2ex]
\multicolumn{7}{c}{$\chi$$^2$ = 3.854, $R_p$ = 5.42\%, $R_{wp}$ = 6.83\% } \\
\hline
\end{tabular}
\end{table}

\section{Results and Discussion}
\subsection{Crystal structure} 

The crystalline structure of Na$_2$Ni$_2$TeO$_6$ can be viewed as a 2D honeycomb network that is composed of edge-sharing NiO$_6$ octahedra with TeO$_6$ in each honeycomb center, and the Na monolayer is sandwiched between two (Ni/Te)-O layers with mirrored orientations along the $a$-axis in each unit cell, as shown in Fig.~\ref{fig-crystru}. Based on the X-ray diffraction results, Evstigneeva \textit{et al.} indexed the diffraction pattern with a space group of $P6_3/mcm$ using three crystallographic sites of Na1 at $6g$  (0.35, 0, $\frac{1}{4}$), Na2 at $4c$ ($\frac{1}{3}$, $\frac{2}{3}$, $\frac{1}{4}$), and Na3 at $2a$ (0, 0, $\frac{1}{4}$), with corresponding occupancies of 44\%, 21\%, and 25\%, respectively.~\cite{Evstigneeva2011}  In this study, neutron diffraction was applied to elucidate the Na ion positions and their corresponding occupancies more precisely. The high-resolution neutron powder diffraction (NPD) data were refined with the Rietveld method using the General Structure Analysis System (GSAS) program.\cite{Larson2004,Rietveld1969}   

\begin{figure*}
\begin{center}
\includegraphics[width=6.5in]{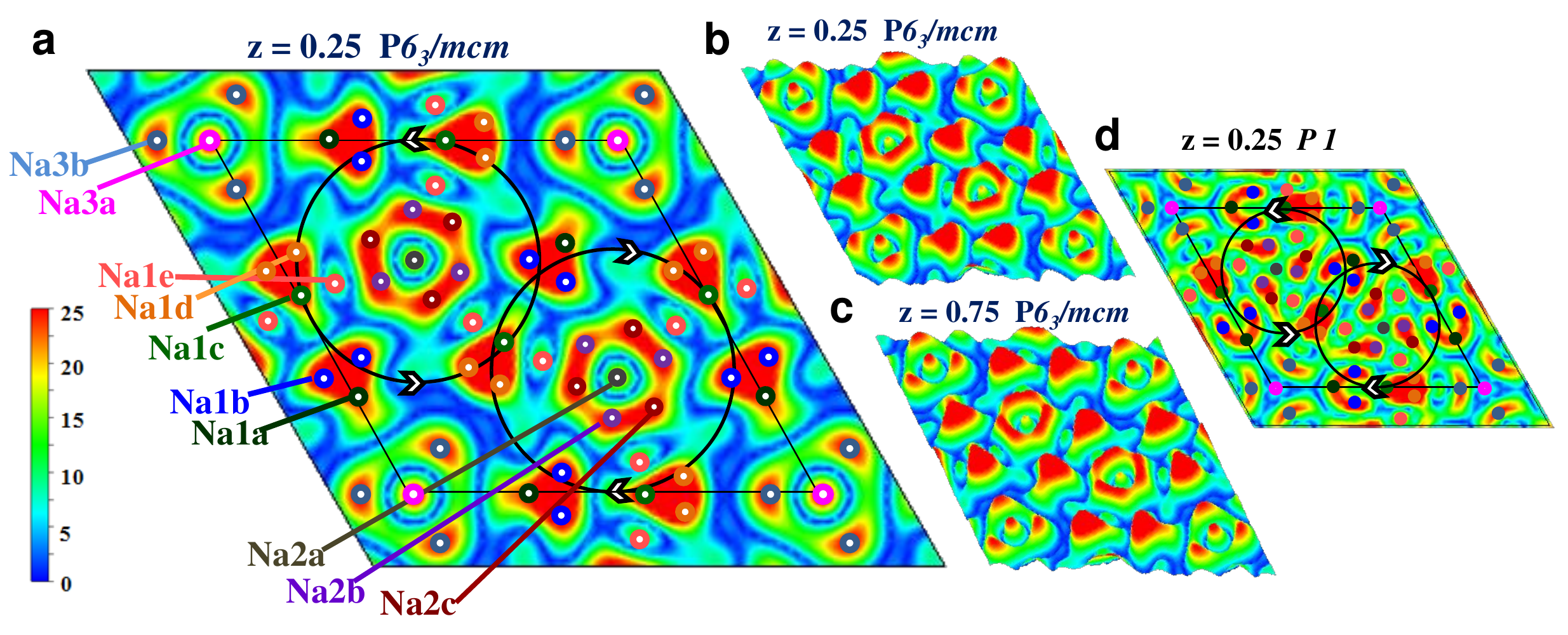}
\end{center}
\caption{\label{fig-NDmap}(color online) (a) The nuclear density map for Na$_2$Ni$_2$TeO$_6$ at z = 0.25 is overlaid with Na sites obtained from the iFT-assisted NPD data refinement, which indicates additional Na1-quintuplet, Na2-triplet, and Na3-doublet splittings are required for the space group $P6_3/mcm$, and the corresponding site occupancies are summarized in Table I. The corresponding 3D contour maps at layers (b) z = 0.25 and (c) z = 0.75 are compared. The nuclear densities of Na1a-Na1d sites are shown with opposite handedness surrounding the Na2a center between the neighboring Na-layers also. (d) Similar handedness can also be identified with space group $P1$ without any presumed symmetry in the hexagonal system, as shown for the z = 0.25 plane via iFT-assisted data analysis using the same NPD data.  }
\end{figure*}

The refinements were performed assuming a hexagonal crystal structure with space group $P6_3/mcm$. Figure~\ref{fig-NPD} displays the observed (crosses) and calculated (solid lines) diffraction patterns taken at 300 K, with differences plotted at the bottom. The nuclear density of the Na layer at z = 0.25 is represented with a map that takes a 0.2 $\AA$ thick cut from the 3D real space view, which is re-constructed from the diffraction data via an inverse Fourier transform (iFT), as  displayed in Fig.~\ref{fig-NDmap}(a)-(c).  It is noted that the original assignments of the Na1, Na2 and Na3 sites do not match the iFT-assisted real space crystal structure of the Na layer perfectly. A better site assignment should reflect the additional Na sites that are revealed by the iFT technique, as indicated by the quintuplet splitting of Na1(a-e), the triplet splitting of Na2(a-c), and the doublet splitting of Na3(a-b). All corresponding occupancies are summarized in the Table I.  The newly assigned Na sites are overlaid to match the nuclear density mapping more satisfactorily, as shown in Fig.~\ref{fig-NDmap}(a). 

It must be noted that these newly identified Na sites from the iFT-assisted NPD data refinement do not offer additional 3D symmetry breaking from the original space group $P6_3/mcm$. In addition, similar iFT-assisted X-ray diffraction data analysis showing electron density distribution (inset of Fig.~\ref{fig-crystru}(b)) is not sufficiently sensitive to resolve the Na sites of low occupancy, which makes current iFT-assisted NPD data analysis a unique and necessary method for the accurate site refinement of the temperature and potential field-sensitive diffusive Na ions. \\

\begin{figure*}
\begin{center}
\includegraphics[width=5.5in]{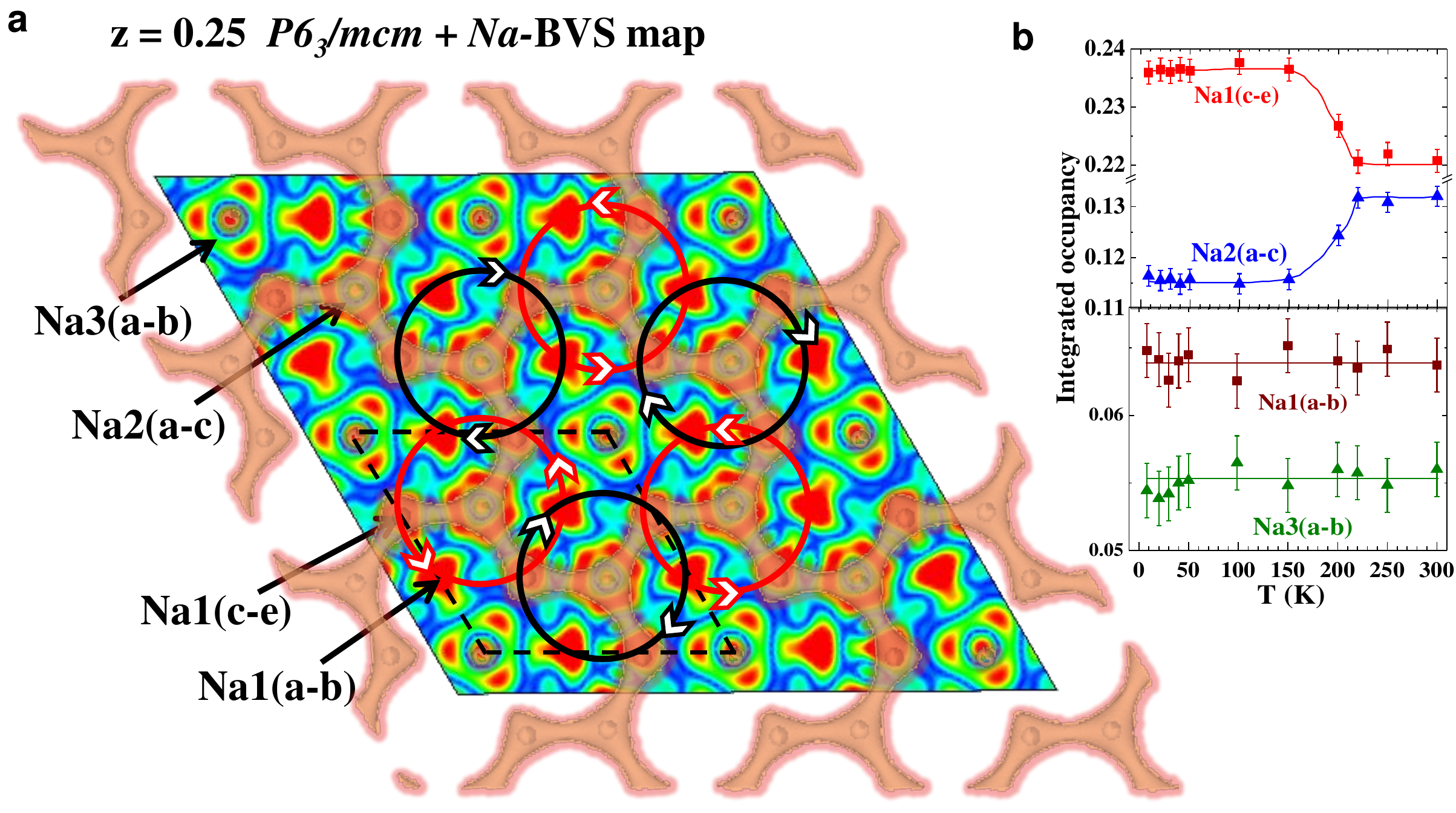}
\end{center}
\caption{\label{fig-diffusion} (a) The nuclear density map of the z=0.25 layer cut for Na$_2$Ni$_2$TeO$_6$ at 300 K, obtained using iFT-assisted NPD data via space group $P6_3/mcm$, where the calculated BVS iso-surface (in brown color) of Na ions is overlaid on top. (b) The temperature dependence of the integrated Na1-Na2-Na3 site occupancies. Solid lines are a guide to the eye. } 
\end{figure*}

\begin{figure*}
\begin{center}
\includegraphics[width=5.5in]{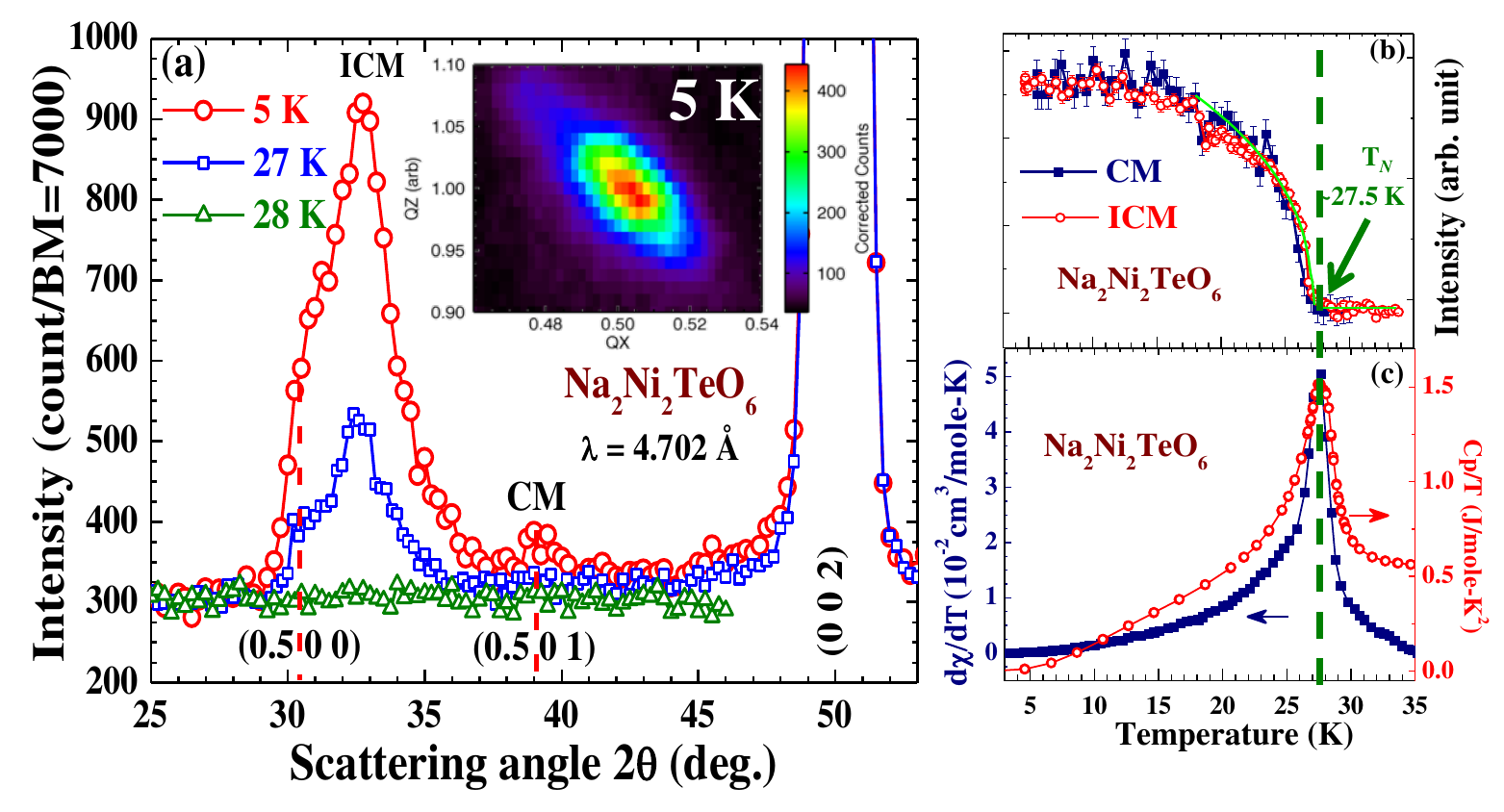}
\end{center}
\caption{\label{fig-OrderPara}(a) CM and ICM AFM peaks observed in the powder sample of Na$_2$Ni$_2$TeO$_6$ below 28 K.  (b) Thermal variations of ICM and CM peaks indicate an onsets of $T_N$$\sim$27.5 K. (c) The d$\chi$/dT and C$_p$/T plots also indicate an anomaly near $\sim$27.5 K.   }
\end{figure*}

\subsection{Sodium distribution and diffusion}

 Following the iFT-assisted NPD data analysis, the experimental nuclear density mapping revealed a circular pattern of well-defined handedness in the Na layer, as illustrated in Fig.~\ref{fig-NDmap}(a)-(c).  Chiral circular pattern can be identified from the nuclear density distribution of the Na1(a-e) sites surrounding the Na2a center in circles showing alternating handedness (counter-clockwise and clockwise).  The observed 2D chiral pattern has also been confirmed by the electron density mapping using X-ray diffraction with the similar iFT-assisted data analysis [inset of Fig.~\ref{fig-crystru}(b)], wherein the electron density (not the nuclear density) distribution of the unresolved Na1 sites indeed shows an in-plane (2D) chiral pattern similar to that identified by the iFT-assisted NPD data [Fig.~\ref{fig-NDmap}(a)]. Such chiral pattern in 2D does not break the 3D symmetry, but its role in 2D symmetry cannot be dismissed completely. For example, such unique local chiral element is found in chiral molecules distributed in 2D. \cite{Elemans2009} The unique chiral element (or named chiral asymmetry) would reduce the 2D symmetry from \textit{p3m1} (No.14) to \textit{p3} (No.13) within the classification of seventeen 2D (Wallpaper) space groups.  
 
 Although a unique potential field reversal is expected between the (Ni/Te)-O layers following the symmetry operation with the space group $P6_3/mcm$, such an in-plane chiral pattern cannot be assigned  unambiguously with the parameters of anisotropic B-factor alone, as indicated by the B-factors listed in Table I. It is noted that distinct chiral pattern can also be identified by applying the iFT-assisted NPD data analysis using the primitive space group $P1$ (No.1) without adding presumed symmetry operation for a hexagonal system, as shown in Fig.~\ref{fig-NDmap}(d), i.e., the observed chirality is intrinsic and not an artifact generated by the choice of space group used in the iFT operation. 

In view of Na$_2$Ni$_2$TeO$_6$ as a layered compound with Na ions intercalated in the van der Waals (vdW) gap, the nuclear density distribution shown in Fig.~\ref{fig-NDmap} could be interpreted as the time-averaged Na ion distribution influenced by the electric potential field, and the potential field is constructed by the neighboring (Ni/Te)-O layers under thermal fluctuation.  This assumption has been supported by the calculated Na ion diffusion path from the molecular dynamic simulations performed earlier.\cite{Sau2015} The iso-surface of the Na bond valence sum (Na-BVS) map shown in Fig.~\ref{fig-diffusion} is calculated from the difference between the experimentally calculated Na-BVS and the ideal valence of +1($\pm$0.2) for the Na ion.\cite{Sale2012} The BVS iso-surface network is continuous across the Na1(c-e) and Na2(a-c) multiplet sites within Na-layer, but leaves out the Na1(a-b) and Na3(a-b) sites completely, as shown in Fig.~\ref{fig-diffusion}(a), which suggests that a Na-ion diffusion path in 2D is established following the Na1(c-e)-Na2(a-c) sites in a circular manner.  A similar Na diffusion path has also been reported earlier in Na$_{0.7}$CoO$_2$ and Na$_3$[Ti$_2$P$_2$O$_{10}$F] via the mean-square displacements and thermal ellipsoid calculations respectively.\cite{Medarde2013, Ma2014}  


It is instructive to examine the Na distribution as a function of temperature for the Na site occupancies in detail, as shown in Fig.~\ref{fig-diffusion}(b). Na1(c-e) shows nearly twice the integrated occupancy than that of Na2(a-c) at 300 K, but there is an additional weight shift of approximately 2$\%$ of the occupancy from Na2(a-c) to Na1(c-e) between $\sim$220-150 K.  On the other hand, the occupancies of Na1(a-b) and Na3(a-b) remain low and constant in the same temperature range.  The actual Na diffusion path is thus hinted by the gradual weight shift of the occupancy between Na1(c-e) and Na2(a-c) experimentally, in agreement with the prediction of BVS calculation.  In addition, it is suggested that Na diffusion occurs above $\simeq$220 K but falls to the relative ground state below $\simeq$150 K, which is consistent to the freezing phenomenon of Na observed in the layered Na$_x$CoO$_2$ below $\sim$150 K via $^{23}$Na NMR.\cite{Weller2009, Schulze2008} 

Since the Na1(c-e)$\leftrightarrow$Na2(a-c) diffusion path (Fig.~\ref{fig-diffusion}) forms closed loops in 2D close packing, a purely rotational diffusion is likely.  In order to introduce an effective translational diffusion, the possible purely rotational diffusion can be avoided by coupling the rotation to an orientational diffusion, as demonstrated by the chiral diffusion for a system of rotary nanomotors.\cite{Nourhani2013}  The revealed chiral circular pattern in 2D must help on making Na$_2$Ni$_2$TeO$_6$ a compound of high ionic conductivity.

\subsection{Magnetic phase transitions} 

 The AFM phase transition for  Na$_2$Ni$_2$TeO$_6$ was verified with elastic neutron scattering using both powder and single crystal samples through the trace of superlattice peaks as a function of the temperature, as shown in Fig.~\ref{fig-OrderPara}(a). At T=5 K, a commensurate (CM) superlattice peak at (0.5 0 1) and a broad incommensurate (ICM) superlattice peak near $\textbf{k}$=(0.47 0.44 0.28) can be identified in the powder sample. While the observed ICM peak has a shoulder close to a possible CM peak of (0.5 0 0) and the ICM peak cannot be deconvoluted with confidence due to the limited instrumental resolution, the peak intensities are plotted as a function of temperature to show that both CM and ICM peaks have same onset at T$_N$ of $\sim$27.5 K, as shown in Fig.~\ref{fig-OrderPara}(b), which has also been identified by the distinct cusp shown in the $d\chi/dT$ and $C_p$/T plots [see Fig.~\ref{fig-OrderPara}(c)]. Our preliminary single crystal studies indicated that the T$_N$ is extremely sensitive to the excess Na content, e.g., T$_N$$\sim$22K has been identified in a single crystal sample of Na content $\sim$2.16.\cite{Karna} A separate spin-polarized neutron experiment is required to elucidate the relationship among the Na content and ICM/CM spin structures further. 
 
We have tentatively explored the CM superlattice peak at (0.5 0 1) using a separate single crystal sample, as shown by the H-L contour plot of the (0.5 0 1) peak intensity [inset of Fig.~\ref{fig-OrderPara}(a)] at 5 K.  The critical exponent of $\beta$ = 0.202 fitted from the powder sample ICM peak intensity with (1-T/T$_N$)$^{2\beta}$ [Fig.~\ref{fig-OrderPara}(b)] suggests the 2D nature of the observed phase transition, which is consistent with the evolution of a short-range exchange coupling prior to the 3D long-range spin ordering, as also revealed by the broad $\chi$(T) peak near $\sim$30-40 K  right above T$_N$ (lower inset of Fig.~\ref{fig-lattice}).  

\subsection{Magnetic structure calculations} 

\begin{figure*}
\begin{center}
\includegraphics[width=5.5in]{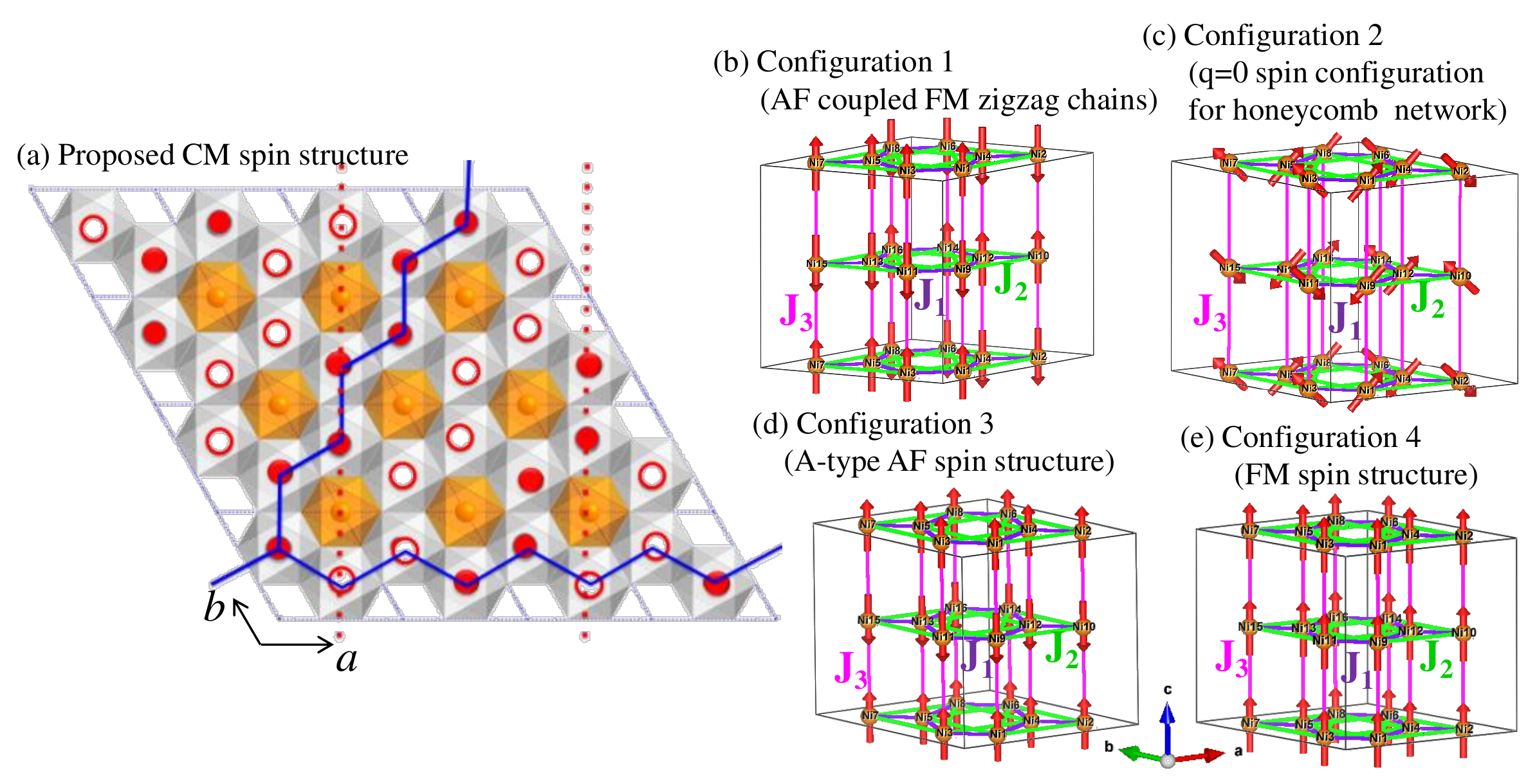}
\end{center}
\caption{\label{fig-SpinStru}(color online)(a) The proposed commensurate AFM spin structure in 2D, where solid and empty circles represent spins up and down. Zigzag chains with doubled periodicity along the $a$-direction is indicated by the dashed lines, which can also be viewed as antiferromagnetically coupled FM zigzag chains extended along the (110) direction. (b)-(e) Four possible spin configurations for the theoretical ground state calculation. The proposed spin configuration shown in (b) is found to be identical and can be described equivalently using spin dimers in linear chain along the $a$- or $b$-direction after a 60$^\circ$ axis transformation within the $ab$-plane, as shown in (a).   } 
\end{figure*}

Judging from the $\chi_{\|c}$(T$<$T$_N$) drop to indicate the on-site spin anisotropy along the $c$-direction,\cite{Sankar2014}  we have considered four possible magnetic configurations of the ground state to evaluate the three spin exchange parameters based on the first-principle density functional theory with a generalized gradient approximation (GGA),\cite{Perdew1996} including an antiferromagnetically coupled FM zigzag chain structure (AF1), a $q$=0 spin configuration for each honeycomb subnetwork (AF2), an A-type AFM spin structure similar to that found in Na$_{0.82}$CoO$_2$ (AF3),\cite{Bayrakci2005} and a FM spin structure were calculated, as illustrated in Fig.~\ref{fig-SpinStru} (b)-(e). The explicit relations between the energy and exchange coupling parameters for the proposed configurations are: $E_{AF1}$ = $E_0$ - 8$J_1$ + 16$J_2$ + 16$J_3$, $E_{AF2}$ = $E_0$ + 8$J_1$ - 16$J_2$ + 16$J_3$, $E_{AF3}$ = $E_0$ - 24$J_1$ - 48$J_2$ + 16$J_3$, and $E_{FM}$ = $E_0$ - 24$J_1$ - 48$J_2$ - 16$J_3$. Considering the in-plane nearest neighbor couplings of $J_1$ (Ni-O-Ni) through the superexchange route and the next nearest neighbor coupling $J_2$ (Ni-O...O-Ni) through the super-superexchange route [Fig.~\ref{fig-crystru}(b)], and an inter-plane coupling of $J_3$, the magnetic ground state is calculated to be the spin configuration of the antiferromagnetically coupled FM zigzag chains (AF1), as illustrated in  Fig.~\ref{fig-SpinStru}(a). The calculated values for the ground state are found to be $E_0$ = -451.37 eV/unit cell, $J_1$ = 0.15 meV, $J_2$ = -1.16 meV, and $J_3$ = -0.08 meV.  As a check for the validity of calculations, a mean-field estimate of Weiss temperature $\Theta$ based on the calculated exchange coupling parameters gives $\Theta$$\sim$$-$26 K, being in good agreement with the experimental value of $-$32 K obtained from the Curie-Weiss law fitting.\cite{Anderson}  

It is noted that the definition of a Ni S=1 spin chain in the 2D honeycomb network is ambiguous when three equivalent zigzag chains can be defined along the (1 0 0), (0 1 0) and (1 1 0) directions of a hexagonal system.  According to the experimental and theoretical informations collected to date, including the identification of the AFM peak ($\frac{1}{2}$ 0 1), the confirmed 3D AFM spin ordering, the spin anisotropy along the $c$-direction,~\cite{Sankar2014} and the theoretically calculated ground state configurations [Fig.~\ref{fig-SpinStru}(b)-(e)] of dominant next nearest neighbor AFM coupling $J_2$ with weak nearest neighbor FM coupling $J_1$, a commensurate AFM spin structure is proposed and shown in Fig.~\ref{fig-SpinStru}(a). This proposed spin structure satisfies all of the experimental and theoretical conditions for the confirmed AFM long-range spin ordering, with a doubled lattice size along the equivalent $a$- and $b$-directions of the hexagonal lattice.  
 
Although the AFM long-range spin ordering is expected to coincide with the lattice size doubling along the $c$-direction of opposite handedness in the Na-layer, strong ICM peaks were identified in the powder sample [Fig.~\ref{fig-OrderPara}(a)], which could have resulted from the incommensurate modulation of the AFM spin arrangement of a 3D helical spin ordering. While the chiral symmetry is closely related to the Coulomb field that is revealed by the nuclear density distribution of the Na-layer, the observed incommensurability could be closely related to the oxygen and/or Na nonstoichiometry.  Preliminary chemical analysis using electron probe microanalysis (EPMA) suggests that the studied powder sample has an oxygen vacancy level about $\sim$0.02 per formula unit.  A careful neutron study on a series of single crystal samples with controlled oxygen and Na nonstoichiometry is expected and in progress. \\

\subsection{Spin-phonon-electronic coupling} 

\begin{figure}
\begin{center}
\includegraphics[width=3.5in]{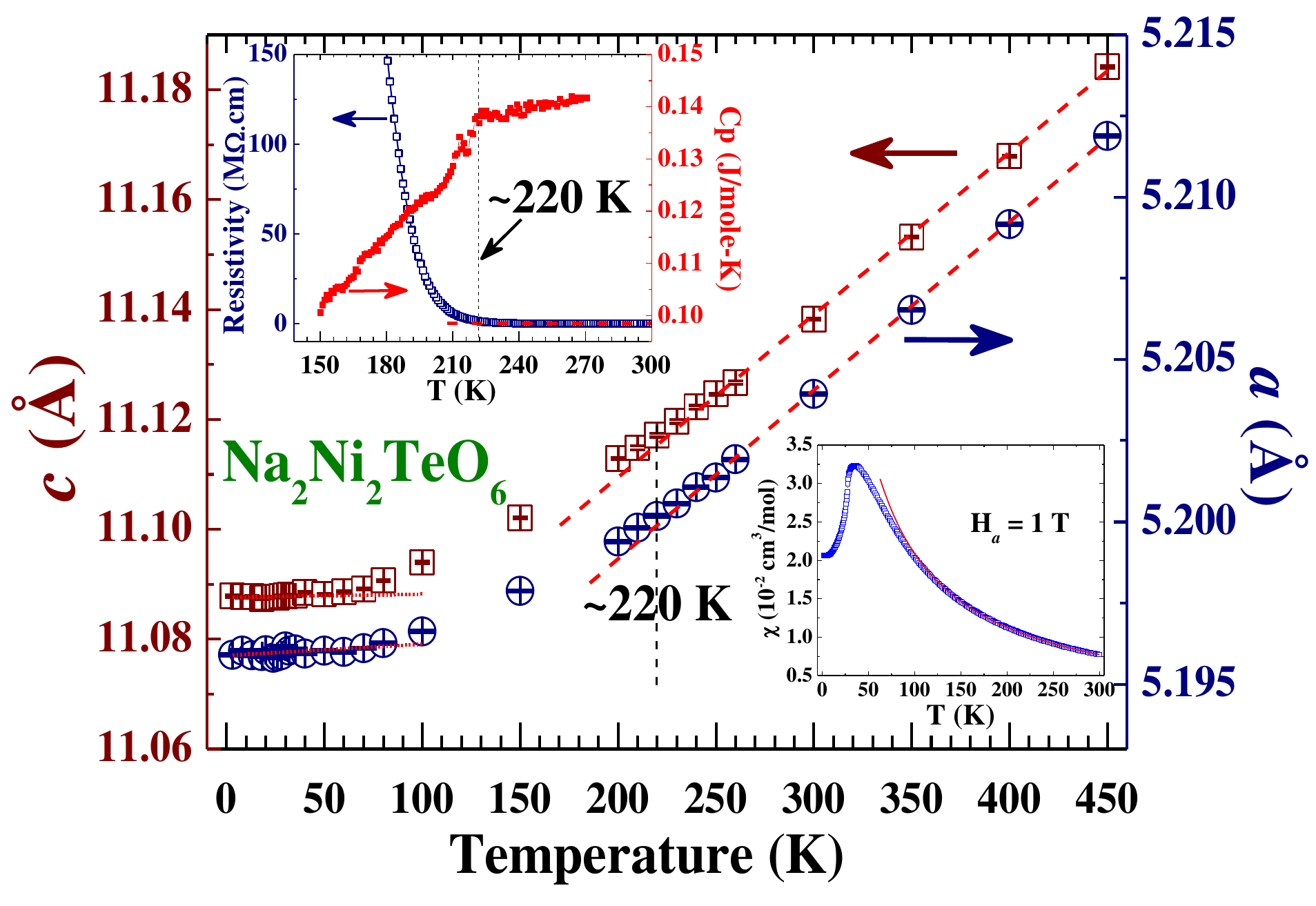}
\end{center}
\caption{\label{fig-lattice} The temperature dependence of the lattice parameters \textit{a} and \textit{c} for Na$_2$Ni$_2$TeO$_6$. The linear temperature dependence is found to be above $\sim$220 K, and the ZTE phenomenon is found to be below $\sim$ 100 K. The upper left inset shows anomalies in the specific heat and resistivity of Na$_2$Ni$_2$TeO$_6$ near $\sim$220 K. The lower right inset shows the temperature dependence of magnetic susceptibilities for powder Na$_2$Ni$_2$TeO$_6$ measured in the applied field of H$_a$ = 1T. The solid line represents the Curie-Weiss law fitting. } 
\end{figure}

\begin{figure}
\begin{center}
\includegraphics[width=2.5in]{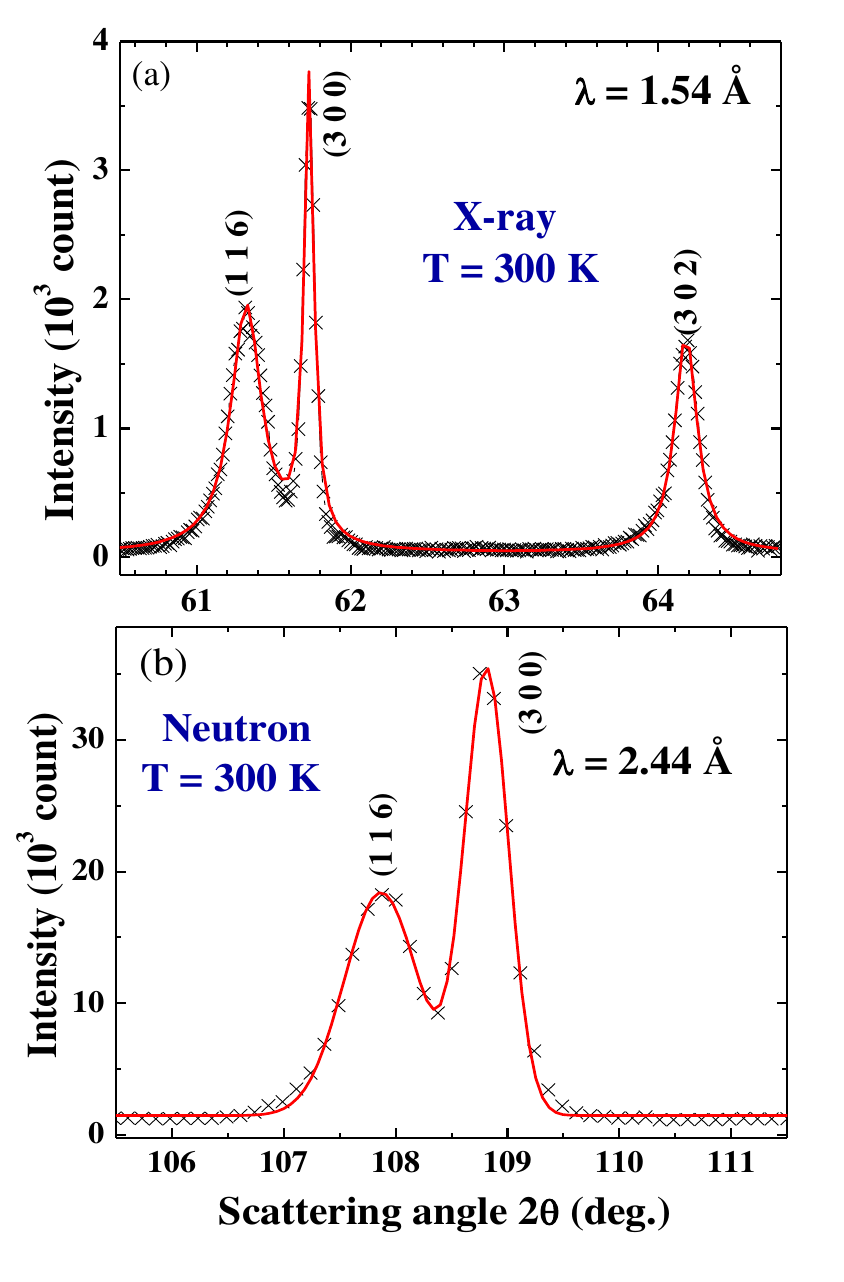}
\end{center}
\caption{\label{fig-strain} A portion of the fitted (a) X-ray and (b) neutron diffraction patterns showing $l$$\neq$0 peaks broadening and $l$=0 peaks with FWHM within the instrumental resolution. All Cu-$K_{\alpha 2}$ for the X-ray has been filtered out.  The calculated corresponding strain is tabulated in Table II.} 
\end{figure}

The lattice parameters $a$ and $c$ for Na$_2$Ni$_2$TeO$_6$ measured as a function of the temperature are shown in Fig.~\ref{fig-lattice}. No structural symmetry change or lattice distortion could be identified from the high-resolution neutron powder diffraction patterns between 3 and 450 K. The thermal expansion was essentially isotropic  above $\sim$220 K, as indicated by the linear temperature dependence. The calculated thermal expansion coefficients are $\alpha_a$ = 9.78$\times$ 10$^{-6}$ K$^{-1}$ along the \textit{a}-axis and $\alpha_c$ = 2.62$\times$ 10$^{-5}$ K$^{-1}$ along the \textit{c}-axis for data above $\sim$220 K. The $\alpha_c$ is approximately 2.68 times higher than $\alpha_a$, which indicates that the inter-layer coupling is considerably weaker than that of the intra-layer coupling, as expected for a layered compound with vdW gaps.

The temperature range for the occurrence of the zero thermal expansion (ZTE) phenomenon below $\sim$100 K is found close to the onset that $\chi$(T) starts to deviate from the Curie-Weiss law (lower right inset of Fig.~\ref{fig-lattice}), indicating both the anharmonic phonon contribution and the short-range spin exchange coupling are required as the precursor to the 3D long-range spin ordering below T$_N$$\sim$27.5 K.  These findings are consistent with a picture of the dominant spin-phonon coupling under reduced thermal fluctuation. The onset of linear thermal expansion near $\sim$220 K was also found near the temperature range where the Na1/Na2 occupancies start to switch weights [Fig.~\ref{fig-diffusion}(b)], which suggests that the subtle in-plane chirality must be closely related to the electron-phonon coupling indicated by the cusp of specific heat C$_p$ and the onset of resistivity increase near $\sim$220 K (upper left inset of Fig.~\ref{fig-lattice}).  Clearly, it is the diffusive nature of the Na ions makes the observation of all these correlated phenomena possible.

\subsection{2D chirality and 3D spin ordering} 

\begin{table}[t]
\caption{Strain calculated for various diffraction peaks shown in the neutron diffraction pattern.  Strain is estimated using Fullprof suite from the integral peak width $\beta$ and the $d$-spacing as $\frac{1}{2}$$\beta$d.   }
\centering
\begin{tabular}{c rrr}
\hline\hline
  HKL                           & ${HG}_o$           & ${HG}_i$    & Strain        
\\ [0.5ex]                 
\hline
(1 0 0)          & 0.4025           & 0.4267        &  Resolution limited \\
(1 0 2)          &  0.4232          & 0.4111       & 21.7251 \\
(3 0 0)          & 0.3141           & 0.3390     & Resolution limited \\
(1 1 6)	       & 0.7227	          & 0.3383	   & 22.0218 \\
(3 0 2)	       &0.4909	          &0.3498	   &11.0855 \\
(1 1 2)	       &0.4116	          &0.3762	    &16.8185 \\
(2 1 1)	        & 0.3426	& 0.3375	      &6.7922 \\[1ex]

\hline
\end{tabular}
\label{tab:afm}
\end{table}

We believe that the hidden chirality in Na layer is resulted from the electric potential gradient created by the neighboring Ni-O layers, which is supported by the detected significant (\textit{h, k, $l$$\neq$0}) peak broadening due to strain calculated using FullProf Suite~\cite{Rodriguez1993} in Table II, in contrast to the narrow (\textit{h, k, $l$=0}) peaks which are within the instrumental resolution (Fig.~\ref{fig-strain}). The different type of bonding between the weaker inter-layer van der Waals interaction and the stronger intra-layer covalent bonding could be responsible for the tolerable anisotropic displacement of oxygen atoms under thermal fluctuation, and the detected strain reflects the required potential gradient for the Na ion diffusion following the Fick's second law.\cite{Shu2008}  Above all, the chirality displayed by the Na ion distribution is hidden in the regular refinement that considers the  statistical atomic position only.

\begin{table}
\caption{ DFT calculations of exchange couplings  $J_1$, $J_2$,$J_3$ for three different types of Na occupancy distribution are compared.}
\scalebox{0.9}{
  \begin{tabular}{c c c c }
    \hline \hline 
   \multicolumn{4}{c}{\textbf{$J_1$/$J_2$/$J_3$}} \\
\hline
  Occupancies                    &100\%Na2               &62.5\%Na1,37.5\%Na2    & 100\%Na1 \\     
    \hline
$J_i$ (meV)          &  0.15/-1.16/-0.08     & -0.11/-1.12/-0.07    &   -0.06/-1.13/-0.05 \\
$J_i$ (K)             &  1.70/-13.47/-0.87    & -1.31/-12.96/-0.83   &   -0.66/-13.08/-0.54 \\

    \hline \hline
 \end{tabular}
}
\label{tab-occu}
\end{table}

In order to test the impact of Na chiral distribution on the spin ordering, DFT calculations of exchange couplings $J_1$-$J_2$-$J_3$ for systems having three different site occupancy assignments are compared, from calculation using the refined occupancies of 62.5$\%$Na1/37.5$\%$Na2 to the fully occupied Na2 and Na1, as shown in Table~\ref{tab-occu}.  It is found that the dominant $J_2$ is not affected significantly by the occupancy change, but the weak $J_1$ and $J_3$ couplings are reduced accordingly. Based on these simulated calculations, we propose that the observed Na-layer 2D chiral distribution via weighted multiplet splitting cannot affect the magnetic couplings significantly.   

While the in-plane antiferromagnetically coupled spin chains [see Fig.~\ref{fig-SpinStru}(a)] may break the mirror symmetry of the honeycomb network via an expected Peierls-like AFM inter-chain coupling, current neutron and X-ray structure analysis failed to detect the expected mirror symmetry breaking, instead, an in-plane chirality was uniquely identified in the Na-layer. It is likely that the 2D chiral distribution pattern identified in the Na-layer may induce the structure relaxation via phonon softening only without breaking the original 3D crystal symmetry. 

\section{Conclusions}

 
In summary, we have demonstrated that the iFT-assisted powder neutron diffraction refinement technique is crucial to provide an accurate site refinement for the diffusive Na ions in the layered material like Na$_2$Ni$_2$TeO$_6$. A chiral distribution pattern hidden in the Na-layer is found to be intimately related to its diffusion behavior and the spin structure of Ni arranged in a honeycomb network.\\
  \\
  \\

\textbf{ Acknowledgements}
FCC acknowledges support provided by MOST-Taiwan under project number 103-2119-M-002-020-MY3. We thank the neutron
group of NSRRC for the support under Grant No. MOST 103-2739-M-213-001-MY3. We acknowledge the support of the ANSTO and NIST for providing the neutron beam time that made the neutron scattering measurements possible. The identification of any commercial product or trade name does not imply endorsement or recommendation by National Institute of Standards and Technology. \\

\end{document}